\pdfoutput=1
\documentclass[useAMS,usenatbib]{mn2e}
\usepackage{graphicx}

\title[The dynamic ejecta of compact object mergers and eccentric collisions]
{The dynamic ejecta of compact object mergers and eccentric collisions}
\author[S. Rosswog]{S. Rosswog$^{1}$\thanks{E-mail:
stephan.rosswog@astro.su.se}\\
$^{1}$The Oskar Klein Centre, Department of Astronomy, AlbaNova, Stockholm University, SE-106 91 Stockholm, Sweden}

\begin{document}


\pagerange{\pageref{firstpage}--\pageref{lastpage}} \pubyear{2012}

\def\paren#1{\left( #1 \right)}
\def\msun{M$_{\odot}$}
\def\Msun{M$_{\odot}$ }
\def\be{\begin{equation}}
\def\ee{\end{equation}}

\maketitle

\label{firstpage}

\begin{abstract}
Compact object mergers eject neutron-rich matter in a number of ways: by the 
dynamical ejection mediated by gravitational torques, as
neutrino-driven winds and probably also a good fraction of the resulting accretion disc finally
becomes unbound by a combination of viscous and nuclear processes. If compact binary 
mergers produce indeed gamma-ray bursts there should also be an interaction region where an ultra-relativistic
outflow interacts with the neutrino-driven wind and produces moderately relativistic ejecta. Each type of ejecta
has different physical properties and therefore plays a different role for nucleosynthesis and for
the electromagnetic transients that go along with compact object encounters. Here we focus on the 
dynamic ejecta and present results for over 30 hydrodynamical simulations of both gravitational 
wave-driven mergers and parabolic encounters as they may occur in globular clusters. We 
find that mergers eject $\sim 1$\% of a solar mass of extremely neutron-rich material. The exact amount
as well as the ejection velocity depends on the involved masses with asymmetric systems ejecting
more material at higher velocities. This material undergoes a robust r-process and both ejecta amount
and abundance pattern are consistent with neutron star mergers being a major source of the "heavy"
($A>130$)  r-process isotopes.
Parabolic collisions, especially those between neutron stars and black holes, eject substantially larger
amounts of mass and therefore cannot occur frequently without overproducing galactic r-process matter. 
We also discuss the electromagnetic transients that are powered by radioactive decays within the 
ejecta (``Macronovae''), and the radio flares that emerge when the ejecta dissipate their large kinetic
energies in the ambient medium.
\end{abstract}


\section{Introduction}
\label{sec:intro}
Even before the discovery of the first neutron star binary PSR 1913+16 \citep{hulse75}
efforts were undertaken to estimate how much mass a compact binary merger would
eject into space.  In a semi-analytic model \cite{lattimer74}
estimated that in a neutron star (ns) black hole (bh) collision $\sim$5\% of the neutron star could become
unbound. By estimating the rates of such encounters they already noted that the accumulated amount 
of ejecta would be roughly comparable to the galactic r-process inventory.
It was only more recently, that the potential of the ejected material to produce electromagnetic (EM) transients 
has been appreciated \citep{li98,kulkarni05,rosswog05a,metzger10b,roberts11,goriely11a,metzger12a,kelley12}. 
EM transients from compact binary mergers are nowadays thought to be instrumental for maximizing the 
science returns from the advanced gravitational wave (GW) detector networks. In fact, EM transients may 
actually provide compelling evidence for the first direct GW detection and they may deliver information 
about the nature of the GW source and its astronomical environment.
%
%
%
%

\begin{figure*}
\centerline{\includegraphics[width=15cm,angle=0]{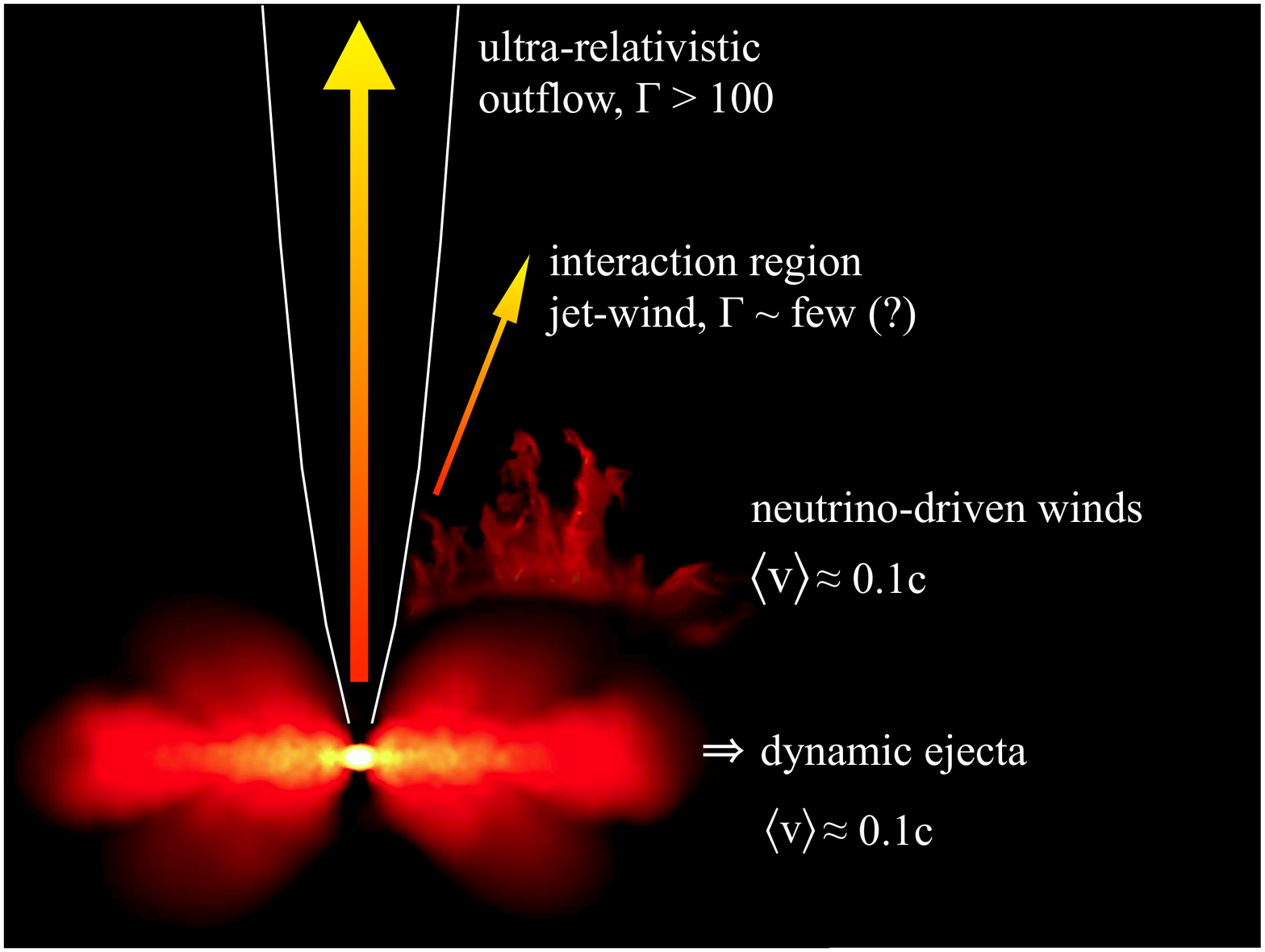} }
\caption{Sketch of the various mass loss mechanisms from a merger remnant ($\Gamma$ denotes
the Lorentz factor). In addition to the illustrated mechanisms there is also a contribution from
the disc-disintegration that is expected to occur at late stages when nucleons recombine into light
nuclei/alpha particles and release a nuclear binding energy comparable to their gravitational binding energy.}
\label{fig:sketch_mass_loss}
\end{figure*}

%
%
The matter required to produce such transients from compact binary mergers can be ejected in a number 
of ways, see Fig.~\ref{fig:sketch_mass_loss}. First, 
there are {\em dynamic ejecta} that are launched immediately at first contact by the interplay between 
hydrodynamics and gravity. This type of ejecta  is the main focus here. In addition, the merger 
remnant emits neutrinos at rates of a few times $10^{53}$ erg/s \citep{ruffert01,rosswog03a,sekiguchi11}
and this neutrino emission has been shown to drive strong baryonic winds \citep{dessart09}. In fact, 
for as long as the central, hypermassive neutron star has not collapsed 
into a black hole, this baryonic wind may actually represent a serious danger for the emergence of 
the ultra-relativistic outflow that is required to produce a gamma-ray burst (GRB). If compact binary 
mergers indeed power short GRBs this neutrino-driven wind should interact near the rotation axis 
with the ultra-relativistic outflow ("jet") producing the GRB. This interaction is expected to 
accelerate matter to  semi-relativistic speeds. 
In addition, at late stages of the disc evolution, viscous heating and/or recombination of free nucleons 
into light nuclei/alpha particles ejects most of what is left from the original accretion disc 
\citep{beloborodov08,metzger09b,lee09}. The ejecta of each of these channels have different properties,  
therefore their role in nucleosynthesis and for the production of EM transients needs to be investigated 
separately for each channel. \\
%
%
Recently, dynamical collisions/high-eccentricity mergers between two compact objects have 
received a fair amount of attention \citep{oleary09,lee10a,rosswog12a,east12,koczis12}. They could, 
for example, be promising GRB central engines and produce
repeated bursts of GWs, but recent  work \cite{rosswog12a} concluded that they can only 
occur at a moderate fraction of the nsns merger rate, otherwise they would cause an overproduction 
of galactic r-process material.\\
Here we discuss the properties of the dynamic ejecta for a large number of simulations of both mergers and 
dynamical collisions.  For both types of encounters we consider  nsns and nsbh systems, the properties of
these simulations are summarized in Tab.~\ref{tab:runs}.

\section{Results}
\label{sec:results}
\subsection{Simulations}
\begin{figure*}
\centerline{\includegraphics[width=11cm,angle=0]{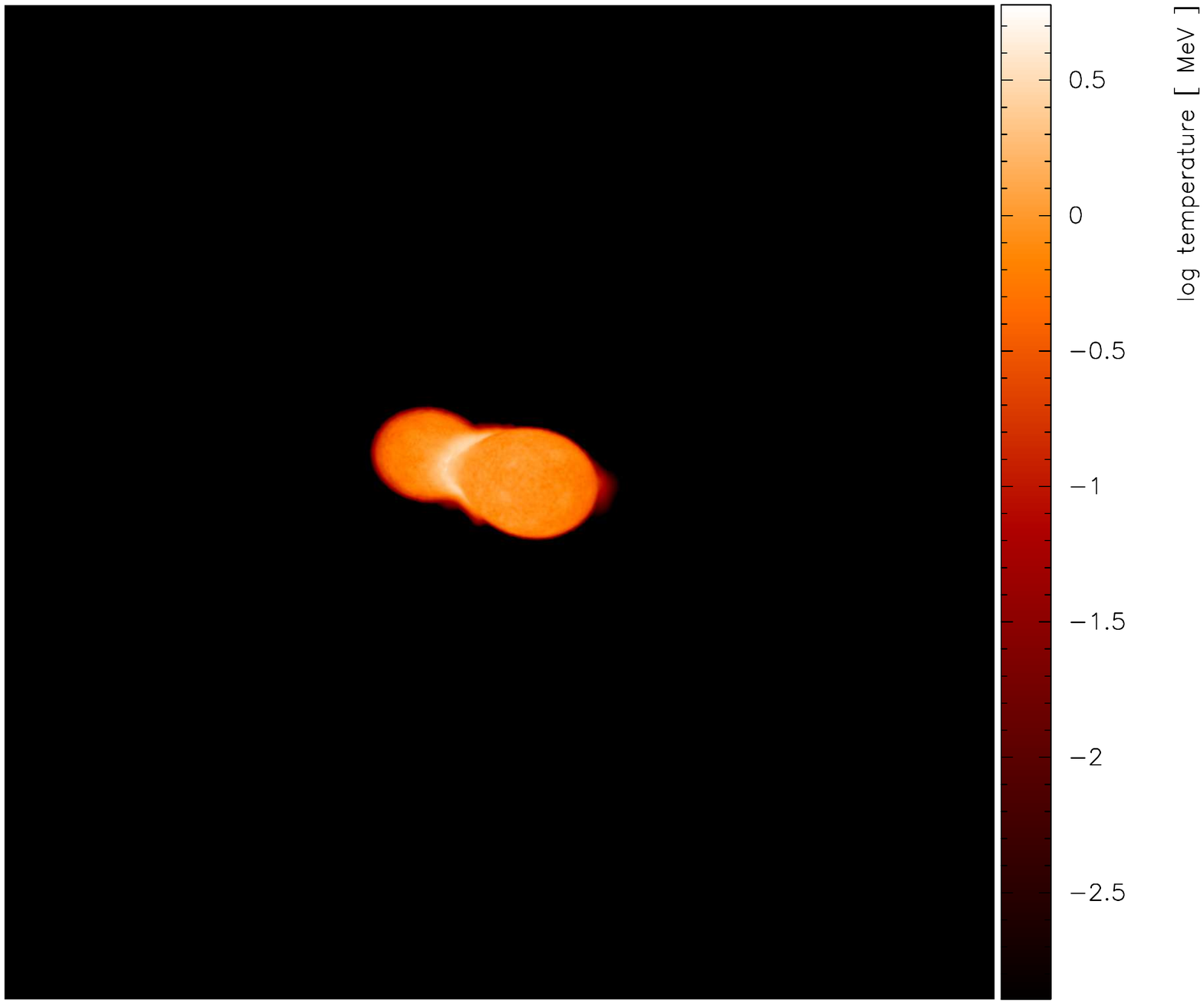} \hspace*{-1.5cm}
                    \includegraphics[width=11cm,angle=0]{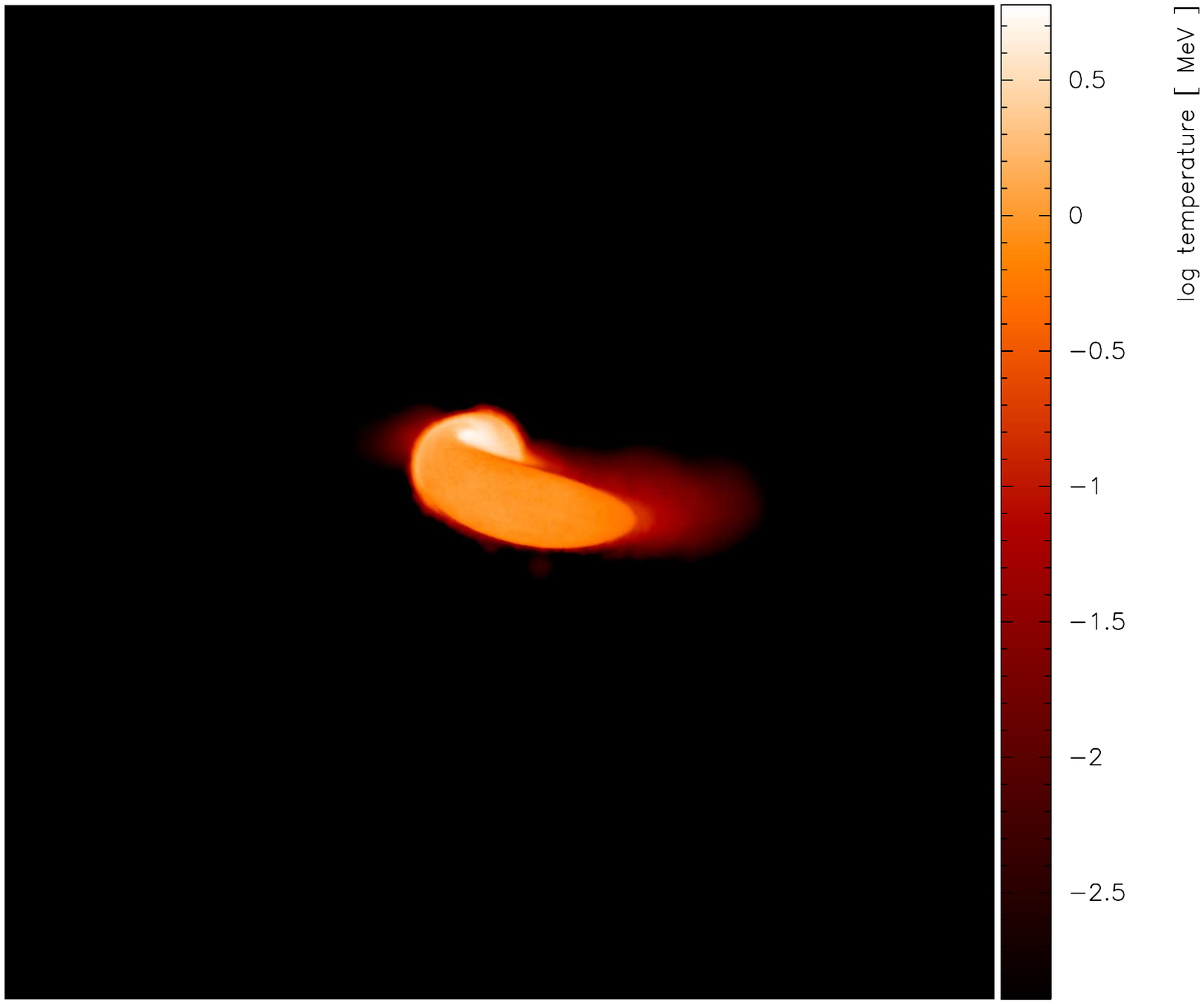} }
\centerline{\includegraphics[width=11cm,angle=0]{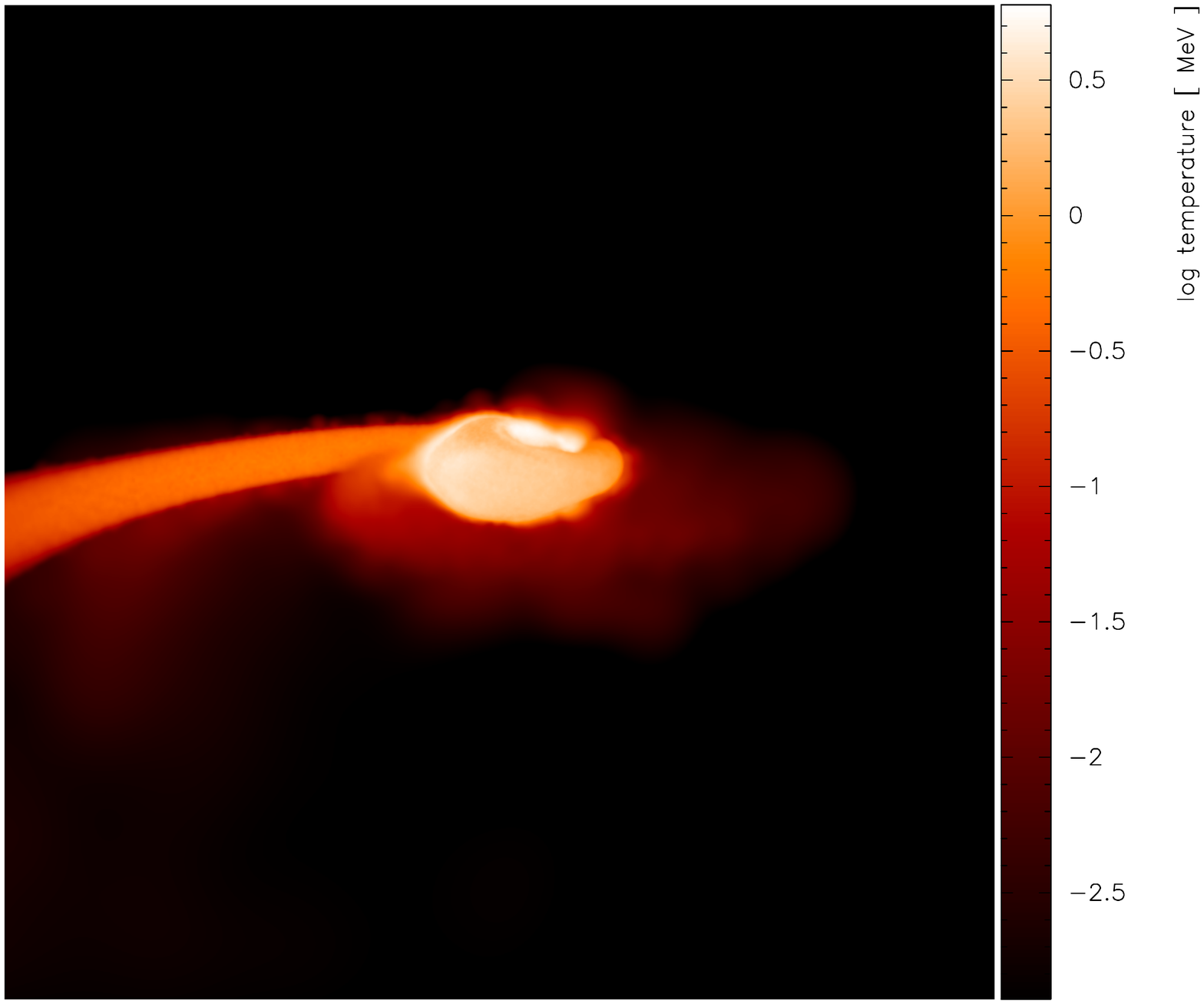} \hspace*{-1.5cm}
                    \includegraphics[width=11cm,angle=0]{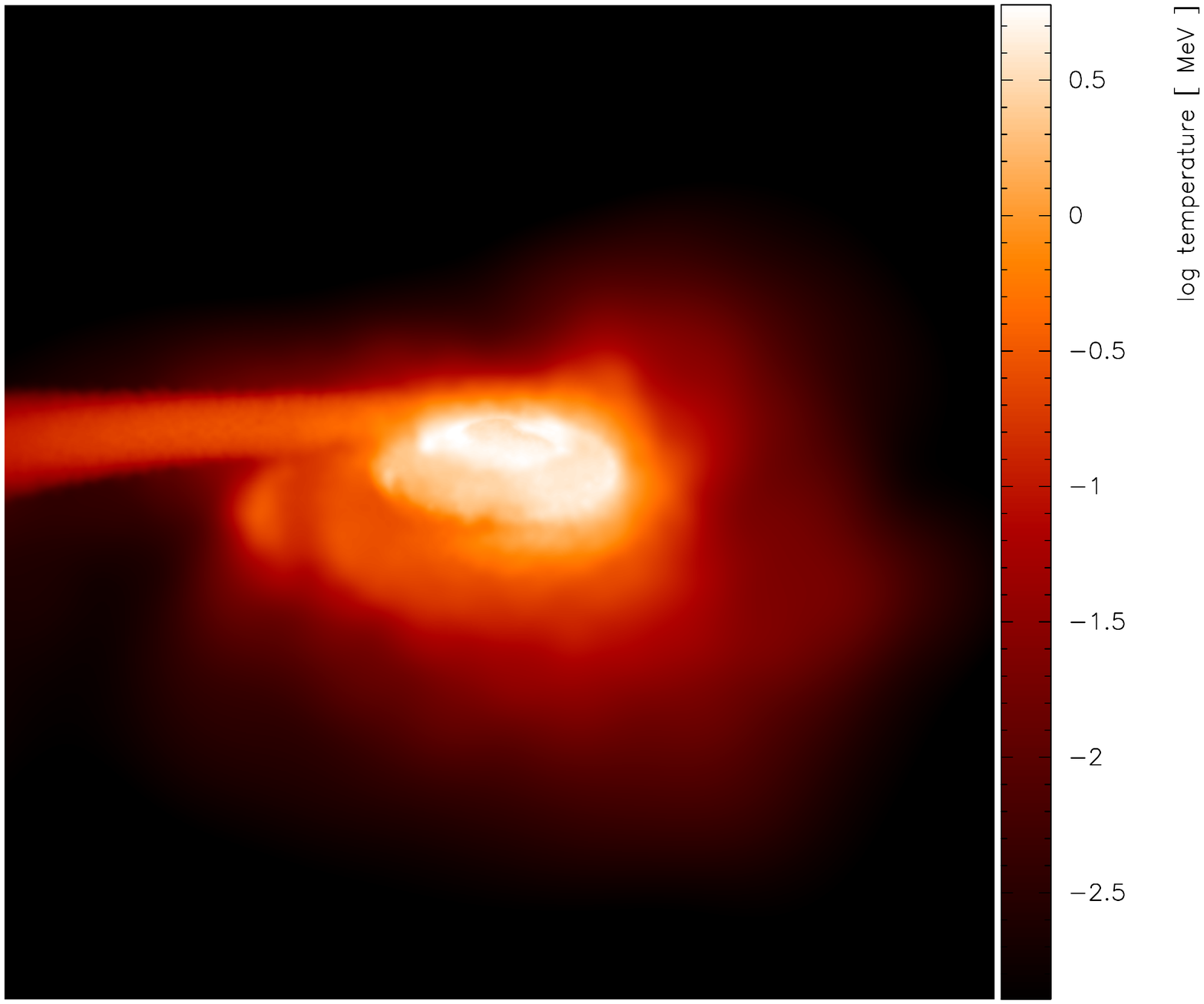} }
\caption{Merger of a 1.3 and a 1.4 \Msun neutron star binary with initial tidal locking. Shown are volume renderings
                 of the temperature at $t=$ 5.04, 6.30, 7.56 and 8.82 ms after simulation start.}                   
\label{fig:nsns_merger}
\end{figure*}
\begin{figure*}
\centerline{\includegraphics[width=11cm,angle=0]{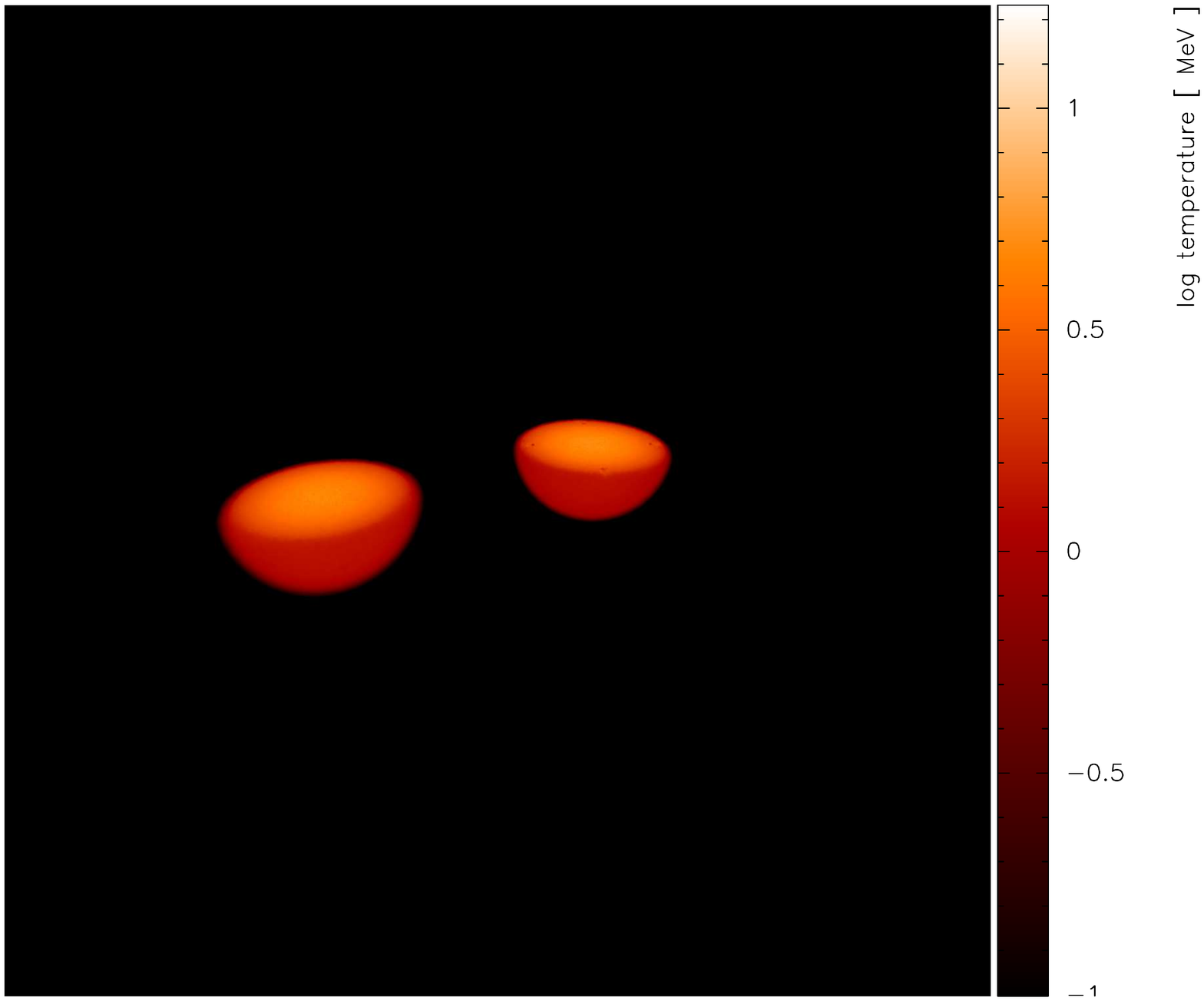} \hspace*{-1.5cm}
                    \includegraphics[width=11cm,angle=0]{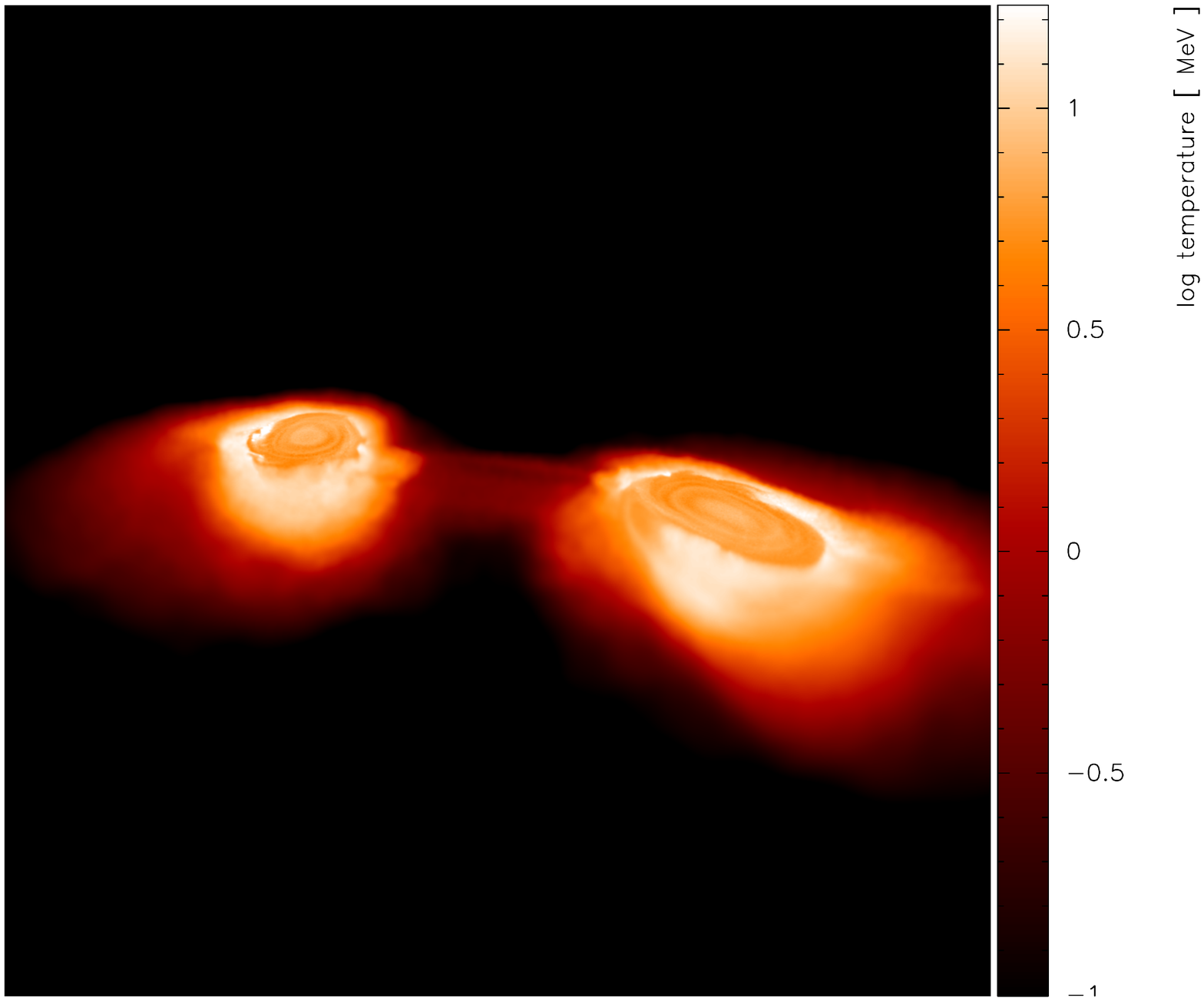} }
\centerline{\includegraphics[width=11cm,angle=0]{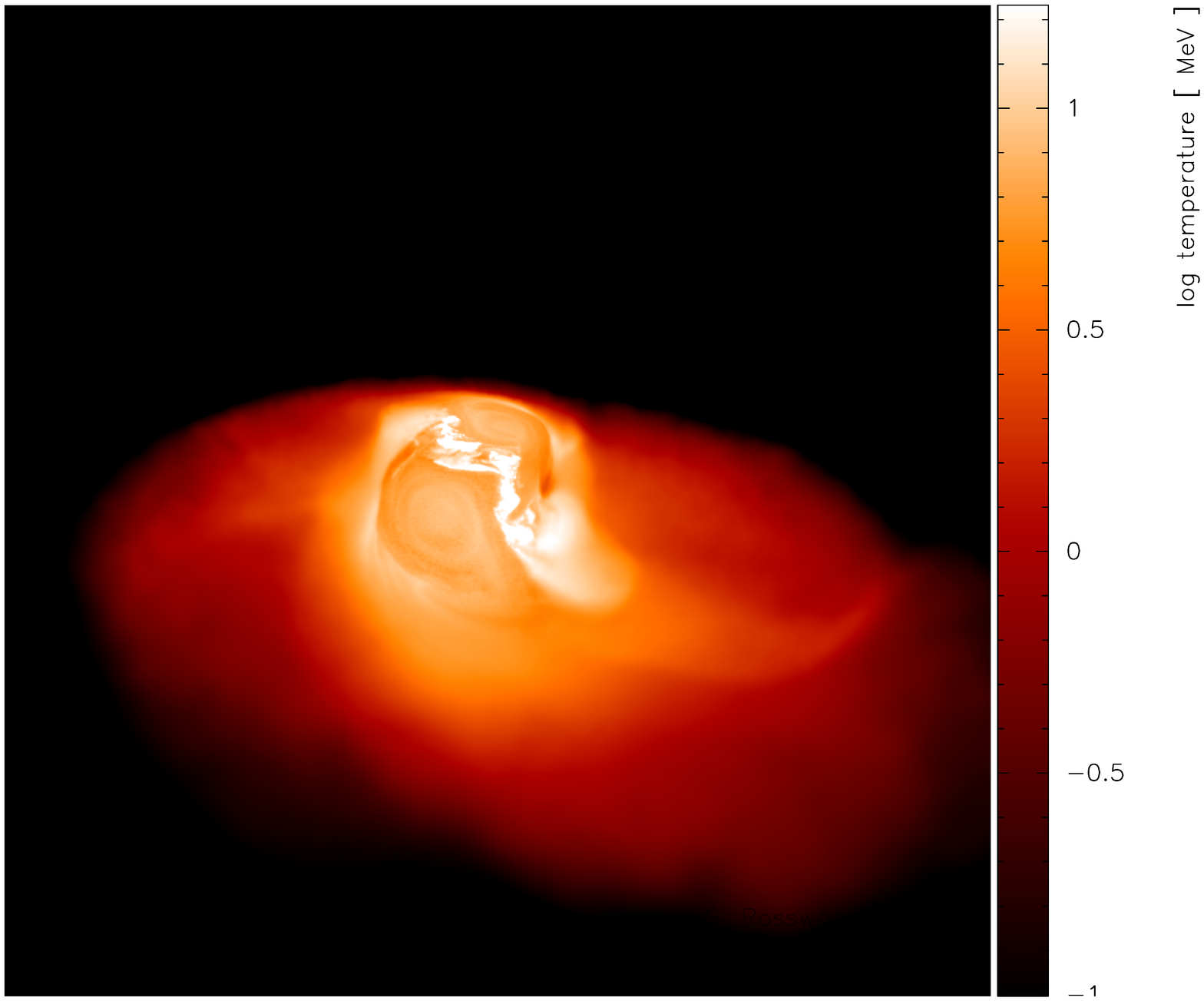} \hspace*{-1.5cm}
                    \includegraphics[width=11cm,angle=0]{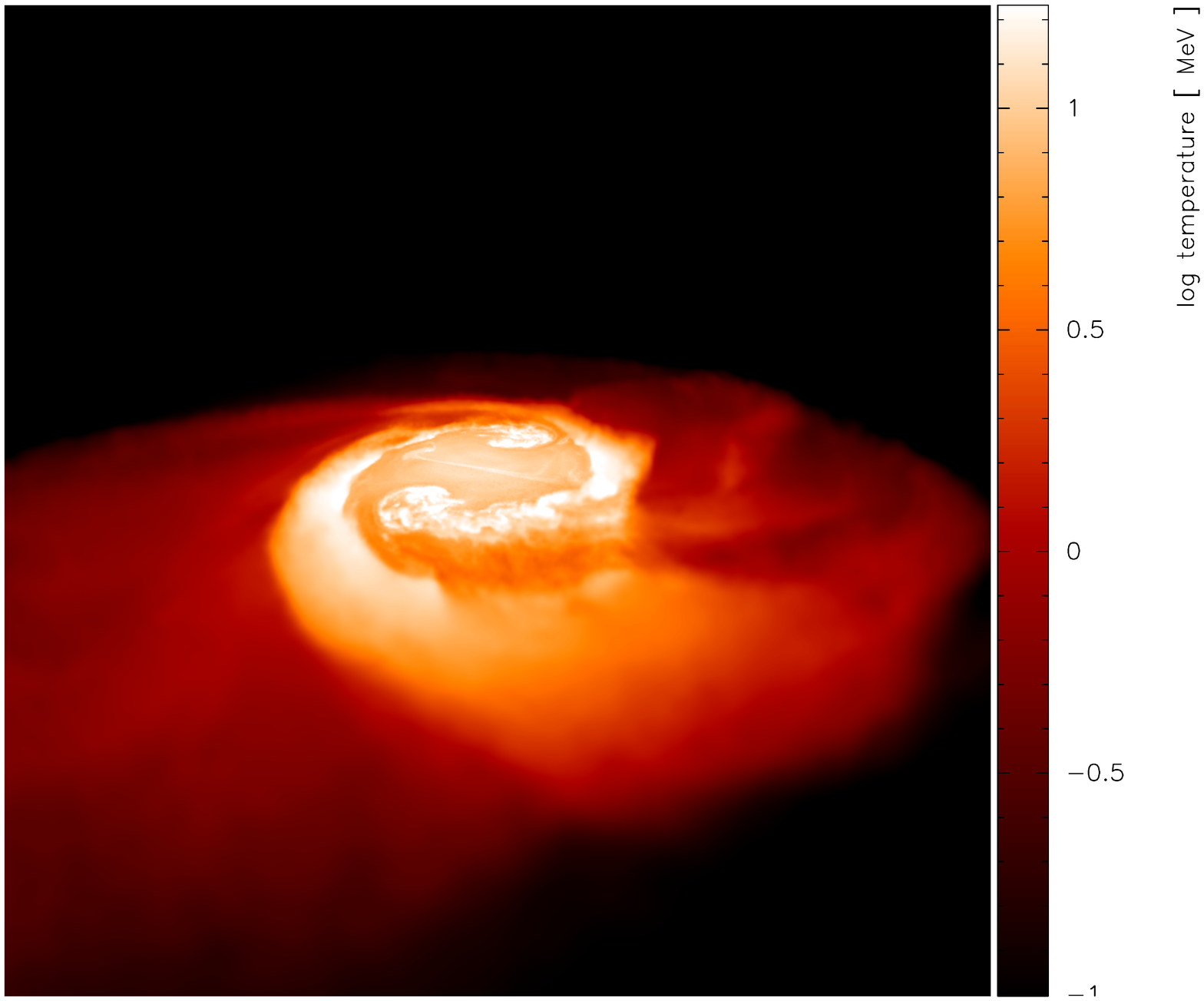} }
\caption{Collision of a 1.3 and a 1.4 \Msun neutron star with an impact strength of $\beta=2$. Shown are volume renderings
                 of the temperature (at $t=$ 1.49, 3.83, 6.27 and 8.32 ms after simulation start), only matter 
                 below the orbital plane is shown.}   
\label{fig:nsns_collision}
\end{figure*}

The presented simulations are performed with 
a 3D Smoothed Particle Hydrodynamics (SPH) code, implementation details can be found in 
the literature \citep{rosswog00,rosswog05a,rosswog07c}, for recent reviews of the SPH 
method consult, for example, \cite{rosswog09b} or \cite{springel10}. The neutron star matter is modeled 
with the Shen et al. equation of state (EOS; \cite{shen98a,shen98b}). We apply an
opacity-dependent, multi-flavor leakage scheme \citep{rosswog03a} to
account for the change of the electron fraction and  the cooling by neutrino emission. 
Black holes are here simply treated as Newtonian point masses with absorbing boundaries at the 
Schwarzschild radius.  For the case of parabolic encounters we parametrize the impact strength
by the ratio $\beta\equiv (R_1+R_2)/R_{\rm per}$, where $R_i$ is the neutron star/Schwarzschild radius
and $R_{\rm per}$ is the separation at pericentre passage. The performed simulations and ejecta properties
are summarized in Tab.~\ref{tab:runs}.\\
We also show two examples to illustrate the hydrodynamic evolution.
Fig.~\ref{fig:nsns_merger} shows volume renderings of the temperature for a neutron star merger case (run 12),
an example of a (substantially more violent) collision between two neutron stars ($\beta=2$; run 27) 
is shown in Fig.~\ref{fig:nsns_collision} (the upper half of matter has been "chopped off" to allow for a view
inside the remnant; the temperature scales are capped to enhance visibility).

\subsection{Dynamic mass loss}
Double neutron star mergers have been known for some time to dynamically eject interesting amounts 
of neutron-rich material \citep{rosswog99}. We consider binary neutron star mergers as the most 
likely type of encounter and run 13 (1.3 and 1.4 \msun, no initial spins) as our reference case. 
Parabolic collisions are interesting, but likely rare encounters whose overall occurrence rate is 
restricted by their large ejecta masses,  see below.
Our reference nsns merger case dynamically ejects $1.4 \times 10^{-2}$ \msun, unequal mass cases 
eject more matter at larger velocities than equal mass cases of the same total mass, see Tab. 1.
For mass ratios $q= m_2/m_1 < 1$ the ejected mass is fit well by 
\be
  m_{ej}(m_1,m_2) =
  (m_1+m_2)\left(A-B\eta-\frac{C}{1+\eta^3/\sigma^3}\right),
  \label{eq:ejecta_fit}
\ee
where $A=0.0125$, $B=0.015$, $C=0.0083$ and $\sigma=0.0056$ \citep{korobkin12a}. Here 
$\eta =1 -4m_1m_2/(m_1+m_2)^2$ is the dimensionless mass asymmetry parameter. 
Collisions of two neutron stars  eject comparable, but slightly larger amounts than double neutron star mergers,
typically a few percent\footnote{Our run 27 has the highest numerical resolution with more than $8 \times 10^6$ 
SPH particles and  is the most expensive of all our simulations, therefore it was only run up to t= 9 ms. 
We consider it likely that the final ejecta amount is larger than the 0.009 \Msun that we measure at the end of
our simulation.}.
We find that collisions between neutron stars and stellar mass black holes, which, due to their larger capture
radius, should dominate over nsns collisions by a factor of $\sim$ 5 \citep{lee10a}, eject significantly 
larger amounts of matter, typically $\sim 0.15$ \msun. Unless equation of state or relativistic gravity effects 
dramatically modify these results, the overall rate of compact object {\em collisions} should therefore be 
seriously constrained, otherwise r-process elements would be substantially overproduced.

\begin{table*}
 \centering
 \begin{minipage}{140mm}
  \caption{Overview over the performed simulations, the superscript $^+$ indicates that the primary 
           is a black hole. Unless otherwise noted, neutron stars have zero initial spin.}
\centerline{\bf Binary mergers}
\vspace*{0.3cm}
\begin{tabular}{@{}rccccccl@{}}
\hline
   Run   &  $m_1$ [\msun] & $m_2$ [\msun] & $N_{\rm SPH}\; [10^6]$ & $t_{\rm end}$ [ms] & $m_{\rm ej}$ [$10^{-2}$\msun]& $\langle v \rangle$ [c]& comment \\
   \hline \\          
   1     &   1.0          & 1.0        & 1.0            &  15.3   & $ 0.76 $& 0.10&\\
   2     &   1.2          & 1.0        & 1.0            &  15.3   & $ 2.5 $ &0.11& \\
   3     &   1.4          & 1.0        & 1.0            &  16.5   & $ 2.9$  & 0.13&\\
   4     &   1.6          & 1.0        & 1.0            &   31.3  & $ 3.1$  & 0.13&\\
   5     &   1.8          & 1.0        & 1.0            &   30.4  & $>1.6 $ &0.13& second. still orbiting\\
   6     &   2.0          & 1.0        & 0.6            &   18.8  & $>2.4$  & 0.16& second. still orbiting\\
\\
   7    &   1.2          & 1.2         & 1.0            &  15.4  & $ 1.7 $ & 0.11& \\
   8    &   1.4          & 1.2         & 1.0            &  13.9  & $ 2.1 $ &  0.12&\\
   9    &   1.6          & 1.2         & 1.0            &  14.8  & $ 3.3 $ &  0.13&\\
   10   &   1.8          & 1.2         & 1.0            &  21.4  & $ 3.4 $ &  0.14&\\
   11   &   2.0          & 1.2         & 0.6            &  15.1  & $>3.0 $ & 0.14 & second. still orbiting\\   
\\
  12    &    1.3        & 1.4         & 2.7            &   20.3 & $5.0 $  & 0.15 &nsns, corot.\\
  13    &    1.3        & 1.4         & 2.7            &   20.3 & $1.4 $  & 0.12&\\   
  14    &   1.4         & 1.4         & 1.0            &  13.4  & $1.3 $  & 0.10&\\
   15   &   1.6         & 1.4         & 1.0            &  12.2  & $2.4$   & 0.12&\\
   16   &   1.8         & 1.4         & 1.0            &  13.1  & $3.8$   & 0.14&\\
   17   &   2.0         & 1.4         & 0.6            &  15.0  & $3.9$   & 0.15&\\   
\\
   18   &   1.6         & 1.6          & 1.0            &  13.2  & $2.0 $ & 0.11&\\
   19   &   1.8         & 1.6          & 1.0            &  13.0  & $1.7 $ & 0.12&\\
   20   &   2.0         & 1.6          & 0.6            &  12.4  & $3.8$  & 0.14&\\
  \\
  21   &   1.8          & 1.8          & 1.0            &  14.0  & $1.5 $ & 0.12&\\
  22   &   2.0          & 1.8          & 0.6            &  11.0  & $2.0 $ & 0.13&\\
  \\
  23   &   2.0          & 2.0          & 0.2            &  21.4  &  $1.2$ & 0.11&\\  
  \\
  24   &   5.0 $^+$   & 1.4       &   0.2             & 138.7  &   $2.4 $ & 0.15& nsbh \\
  25   &  10.0 $^+$   & 1.4      &   0.2            & 139.0  &   $4.9 $ & 0.18 & nsbh \\
  \\
\end{tabular}
\\
\centerline{\bf \hspace*{2.5cm} Parabolic collisions\\}
\vspace*{0.0cm}
\\
\begin{tabular}{@{}rccccccl@{}}
\hline
   Run   &  $m_1$ [\msun] & $m_2$ [\msun] & $N_{\rm SPH}\; [10^6]$ & $t_{\rm end}$ [ms] & $m_{\rm ej}$ [$10^{-2}$\msun]& $\langle v \rangle$ [c]& comment \\
   \hline 
   \\
    26 &  1.4 & 1.3 &   2.7 & 21.2 & $6.0 $ & 0.13&nsns, $\beta=1$   \\ 
    27 &  1.4 & 1.3 &   8.0 & 9.0 & $  0.9 $ & 0.22&nsns, $\beta= 2$  \\
    28 &  1.4 & 1.3 &   2.7 & 13.2 & $  3.0 $ & 0.28&nsns, $\beta= 5$   \\
    \\
    29 &  3.0 $^+$ & 1.3 &   1.3 & 127.5 & $14.2 $ & 0.19&nsbh, $\beta=1$  \\    
    30 &  5.0 $^+$ & 1.3 &   1.3 & 143.6 & $17.2 $ & 0.24&nsbh, $\beta=1$  \\      
    31 &  10.0 $^+$ & 1.3 &   1.3 & 540.3 & $13.4 $ & 0.24&nsbh, $\beta=1$  
     \label{tab:runs}
\end{tabular}
\end{minipage}
\end{table*}

\subsection{Impact on heavy element nucleosynthesis}
Approximately half of the elements heavier than iron are formed by rapid (in comparison 
to $\beta$-decays) neutron capture reactions or ``r-process'' for short. The r-process
production sites, however, are still matters of debate. The r-process elements observed 
in metal-poor stars \citep{sneden08} point to at least two groups of r-process events. 
One occurs relatively frequently and produces predominantly lighter elements from strontium 
to silver \citep{cowan06,honda06}. It may actually be the result of a 
superposition of a number of different sources. The other one is rarer
and produces whenever it occurs the heaviest r-process elements (beyond Ba, $Z=56$) 
in nearly exactly solar proportions. So far, there is no generally accepted explanation 
for the robustness of this unique heavy r-process component.\\
Traditionally supernovae were considered the most likely source of r-process elements, but 
a number of recent investigations has cast doubts over this view (e.g. 
\cite{arcones07,roberts10,fischer10,huedepohl10}). The main contenders of 
supernovae in terms of r-process nucleosynthesis are compact binary mergers of either two 
neutron stars or a neutron star and a stellar-mass black hole 
\citep{lattimer74,lattimer76,eichler89,freiburghaus99b}.\\
The matter that is ejected by double neutron star mergers shows a narrow distribution 
of electron fractions around $Y_e\approx 0.03$. Collisions produce a slightly broader 
but still very neutron-rich distribution, see Fig. 11 in \cite{rosswog12a}.  The 
latter occurs due to the larger temperatures in collisions which allow positron captures 
to increase $Y_e$. We have explored the nucleosynthesis within these ejecta \citep{korobkin12a}
and found  a very robust r-process nucleosynthesis which produces all the 
elements from the second to the third r-process peak in close-to-solar ratios. The extreme neutron 
richness makes the r-process path meander along the neutron drip line and, as a result, 
the final abundance patterns are predominantly determined by nuclear properties rather 
than by those of the merging astrophysical system. 
Consequently, all cases produce essentially identical
abundance patterns, see Fig. 4c in \cite{korobkin12a}. Substantial deviations from this pattern only occur 
for trajectories that have initial $Y_e$-values above $\sim$0.17 (see Fig. 8 in Korobkin et al. 2012). 
Such material, however, is too rare to have a noticeable impact on the  resulting abundance pattern. 
\begin{figure*}
\centerline{\includegraphics[width=10cm,angle=-90]{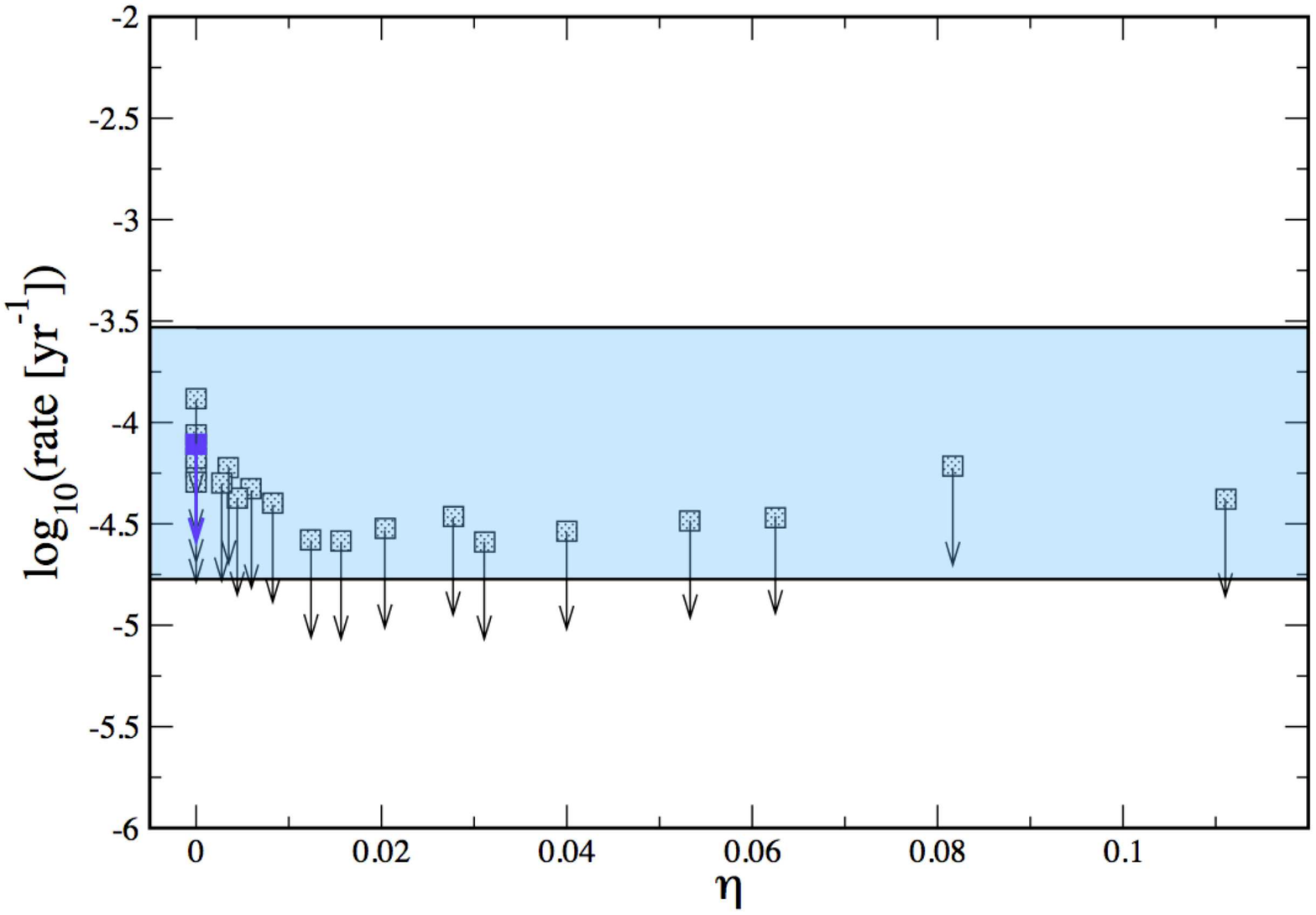}} 
\caption{Relevance for cosmic r-process inventory. Shown is the event rate that is, based on the simulation results, 
required to reproduce the galactic r-process enrichment rate, $\dot{M}_{\rm r,gal}$, for a 
given mass asymmetry parameter $\eta$. For each value of $\eta$ the required rate (square) is calculated as 
$\dot{M}_{\rm r,gal}/m_{\rm ej, sim}$, where $m_{\rm ej, sim}$ is the ejecta mass found in 
the simulations. For comparison, the expected 95\% confidence interval for the nsns merger rate as derived from 
observations (Kalogera et al. 2004) is shown. Within the existing uncertainties, the ejecta masses are consistent
with nsns mergers being major production sites of r-process material.}                 
\label{fig:rates_r_process}  
\end{figure*}    
    
\subsection{Constraints on occurrence rates}
To gauge the possible relevance of nsns mergers for the enrichment of the Cosmos with heavy elements, we 
take the ejecta masses found in the simulations and make the (very strong) assumption that the considered mergers
are the only source of r-process elements. The occurrence rates that are required under these assumptions  to
reproduce the galactic r-process enrichment rate of $\sim 10^{-6}$ \Msun yr$^{-1}$ \citep{qian00} are plotted in 
Fig.~\ref{fig:rates_r_process}. The quantity $\eta$ is the dimensionless mass asymmetry parameter $\eta$ that was also
used in Eq.~(\ref{eq:ejecta_fit}). These required rates are compared to the 95\% confidence interval for the
estimated nsns merger rate that is based on the observed binary neutron star population \citep{kalogera04a}.
Thus, within the existing uncertainties, double neutron star mergers eject enough material to deliver a major
contribution to the Galactic r-process record.\\
Collisions, on the other hand, eject substantially larger amounts. nsbh collisions are expected to occur
five times more frequently than nsns collisions \cite{lee10a}, therefore an average collision ejects 
$\sim (5 \times 0.15 + 0.06)/6 \sim 0.135$ \Msun per event, about an order of magnitude more than the typical
merger case. We can obtain a robust upper limit on the collision rate if we make the extreme assumption 
that collisions are the only producers of r-process and that no other event contributes. Under this extreme
assumption their occurrence rate would be about 10\% of the nsns merger rate. The rate realized in nature 
may actually be well below this value.

\subsection{Electromagnetic transients}
The  electromagnetic transients that go along with a compact binary merger have recently attracted much 
interest \citep{li98,kulkarni05,rosswog05a,metzger10b,roberts11,goriely11a,metzger12a,kelley12}. 
Such transients can provide information on the distance, on the type 
of and the position with respect to the host galaxy, on its metallicity and on the ambient matter densities.
In other words EM transients help to place a compact binary coalescence into an astrophysical context.
After having been operational intermittently during the last decade the LIGO and VIRGO detectors are 
currently being upgraded \citep{abbott09a,sengupta10}. 
By about 2016 they should reach their new design sensitivities which are 10-15 times higher than those 
of the initial instruments, so that the accessible volume increases by more than three orders of magnitude. 
This will push the detection horizons for nsns mergers out to a few hundred Mpc and to nearly a Gpc 
for nsbh mergers \citep{abadie10}. 
The first detections are expected to be near threshold and accompanying EM signals could substantially 
boost the confidence in a candidate event and thus effectively increase the instrument sensitivities 
\citep{kochanek93,hughes03,dalal06,arun09}. Rather than relying on 
accidental coincident EM and GW detections one could increase the detection rate by either following a
GW candidate event by target-of-opportunity searches for EM transients or by scanning through archival data
based on EM triggers (e.g. \cite{kochanek93,mohanty05,mandel10,nakar11a,metzger12a}).\\
\subsubsection{GRBs}
The most luminous expected EM transients  are short Gamma-ray bursts (sGRBs; 
\cite{paczynski86,eichler89,narayan92,piran04,nakar07,lee07}. 
The physical mechanisms behind launching such a burst are still far from being settled, it seems, however, 
that the ultra-relativistic outflows that produce the bursts are collimated into half-opening angles of  
$\sim 5^\circ$ \citep{fong12}, consistent  with theoretical expectations \citep{rosswog03b,aloy05}. 
As discussed previously, the 
emerging neutrino-driven winds pose a serious threat to the emergence of an ultra-relativistic outflow and 
therefore not every  nsns merger may actually be able to produce a sGRB. In other words, the detected sGRB 
rate may possibly be only a small fraction of the true nsns merger rate, $R_{\rm sGRB}= f_b f_p R_{\rm nsns}$, where $f_b$ 
is the beaming fraction and $f_p$ the fraction of nsns mergers that is not choked by baryonic pollution
(e.g. through neutrino-driven winds). 
Cases where a GW signal is detected but no GRB might then be used as trigger on those EM transients that 
are less beamed such such as "orphan afterglows" or macronovae, see below. \\
\subsubsection{Macronovae}
"Macronovae" are radioactively powered transients that emerge from the decaying ejecta of compact object mergers 
\cite{li98}.  In contrast to GRBs, they should be "isotropic" in the sense that they are visible from all
sides, although the ejecta distribution suggests a viewing angle dependence, see Figs. 1 and 2 in Piran et al. 2012.
Macronovae share some similarities with supernovae, in particular, without late-time energy injection from radioactive 
decays they would be hardly detectable at all. The ejecta composition of a macronova is unique and very 
different from any type of supernova. While the latter produce elements up to the iron group near $Z=26$, 
the dynamic ejecta of neutron star mergers consist entirely of r-process elements up to the third peak near $Z  \approx 90$, 
see above, and should thus leave a distinctive imprint on the observable electromagnetic display.\\
Given the expected complexity of the involved physics, the models that exist to date are still rather basic.
They rely on estimates for the ejected mass and its velocity distribution (or alternatively $dm/dv$), which 
can be straightforwardly extracted from hydrodynamics simulations. Via nuclear network calculations along hydrodynamic
trajectories one can extract the nuclear energy injection rate, $\dot{\epsilon}$, which shows little sensitivity  to the
exact details \citep{metzger10b,korobkin12a}.  The latter authors find that 
\begin{equation}
  \dot{\epsilon}(t) = 2 \times 10^{18} \frac{\rm erg}{\rm g s}\left(\frac{1}{2}-\frac{1}{\pi}
                  \arctan{\frac{t-1.3 {\rm s}}{0.11 {\rm s}}}\right)^{1.3}
                  \times\left(\frac{\epsilon_{th}}{0.5}\right)
\end{equation}
provides a good fit to the results of the network calculations. Here $\epsilon_{\rm th}$ is the fraction of energy
that is injected as thermal radiation. The last ingredient of existing models is
an average value for the opacity $\kappa$. The opacity value is crucial since it determines the diffusion 
time which, in turn, sets the time of peak emission. In existing models this value has been taken as 0.1 cm$^2$/g
which is characteristic of the line expansion opacity from iron group elements (e.g. \cite{kasen07}).\\
With these assumptions  one finds that a macronova resulting from a 2 $\times$ 1.4 \Msun nsns merger peaks
after $\approx 0.4$ days with $L_{\rm peak}\approx 5 \times 10^{41}$ erg/s \citep{piran12a}. Nearly all other nsns and nsbh
cases, both mergers and collisions, deliver larger ejecta masses and velocities and therefore produce larger
luminosities (up to $\sim 10^{42}$ erg/s) at slightly later times ($t_{\rm peak}<1$ day) see Fig. 13 in Piran et al. (2012)
for the merger and Fig. 15 in \cite{rosswog12a} for the collision cases. Other groups \citep{metzger10b,goriely11a,roberts11}
find similar results for nsns merger cases. \\
It was recently pointed out \citep{kasen12} that the opacities of the freshly synthesized r-process
material are very poorly known. Most likely, they will be dominated by millions of Doppler-broadened lines from
the lanthanide group ($Z=57 - 71$). Little atomic data is available to date for these elements, but initial calculations 
of Kasen and collaborators suggest that the opacities might be about two orders of magnitude larger than 
those of  iron group elements.  As a result,  the larger diffusion times are expected to lead to lightcurves that 
peak only after several days rather than just a few hours. The effective temperatures will be substantially degraded
so that the bulk of radiation may actually escape in IR rather than optical/UV (optical is suppressed by 
line blanketing). The important topic of macronova transients certainly deserves more efforts in the future.\\
\subsubsection{Radio transients}
It is important to realize that  compact object merger ejecta can contain a kinetic energy that is comparable to a supernova.
For example, the ejecta of a  typical nsns merger (1.3 \& 1.4 \msun) contain $2 \times 10^{50}$ erg. The more extreme, 
but probably rare collision cases can even contain up to $10^{52}$ erg of kinetic energy \citep{rosswog12a}. The 
deceleration of this sub-/mildly relativistic material drives a strong shock into the ambient medium. 
Shocks with similar properties have been observed in late stages of GRB afterglows and in the early phases of some 
supernovae. In both circumstances they produce bright radio emission. The ejecta of compact object encounters are 
sprayed with a distribution of velocities into the surroundings, for typical nsns mergers the average 
velocity is close to 0.1c.  Asymmetric mergers ($q\ne1$)  deliver average velocities up to 0.18c, but
how often these higher-velocity cases occur depends on their not so well-known  mass distribution. 
Collisions yield substantially larger average velocities (up to 0.28c, see Tab. 1) with a high 
velocity tail reaching close to the speed of light. The highest of these velocities, however, need to
be interpreted with care since the simulations are essentially Newtonian. Nevertheless, it is a robust result that
dynamical collisions produce substantially larger ejecta velocities, therefore their radio signals are brighter 
and peak earlier. It needs to be re-iterated, though, that collisions must be rare in comparison to nsns mergers.\\
With assumptions similar to those successfully applied in GRB afterglows, i.e. constant internal energy fractions 
behind the shock in electrons, $\epsilon_e=0.1$, and in magnetic field, $\epsilon_B=0.1$, and electrons being 
accelerated into a  power law distribution  with index $p=2.5$, one can estimate the resulting radio 
emission (for a detailed description see \cite{piran12a}).
At 1.4 GHz the radio signal that emerges from a typical nsns merger at the detection horizon of advanced 
LIGO (300 Mpc) remains on a level of $\sim 50\mu$Jy for several years, provided that it occurs in an 
environment similar to the one of the observed Galactic nsns systems where the density is of order 
1 cm$^{-3}$. Asymmetric $q\ne1$ mergers and in particular collisions with their larger ejecta velocities 
and masses produce brighter and longer lasting radio flares \citep{piran12a,rosswog12a}.
The ambient matter density is a major uncertainty for the detectability, though. Neutron star binary 
systems that receive a kick at birth could easily travel out of the galactic plane and merge where 
densities are substantially lower \citep{fryer99a,bloom99,rosswog03c,belczynski06,fong10}. The transients 
from the dynamic ejecta of such cases would be very hard to detect.\\
Due to their velocities of $\sim 0.1$c, the transients resulting from the dynamic ejecta peak after 
about one year. If present, mildly relativistic outflows would dominate the radio emission at earlier 
times. The physics to account for such outflows is not included in the presented simulations, such 
mildly relativistic outflows are, however, likely to occur. As sketched in Fig.~\ref{fig:sketch_mass_loss}, 
they are expected  at the interface between the ultra-relativistic outflows that trigger the sGRB 
and the neutrino-driven winds. It is a numerical challenge, though, to calculate  their velocity 
structure/Lorentz factor distribution in a reliable way. The dynamics in the interaction region will be 
dominated by fluid instabilities such as Kelvin-Helmholtz where the shortest perturbations grow fastest. 
These perturbations are usually set by the finite numerical resolution length. Nevertheless, since this 
type of outflow may dominate the early radio emission it deserves further studies in the future.

\section{Discussion}
\label{sec:discussion}
We have briefly summarized the ways in which a compact binary encounter ejects neutron-rich matter into
its surroundings. Our focus was on the dynamic ejecta, launched by hydrodynamic effects and
gravitational torques, and on their implications for nucleosynthesis and EM transients going along 
with a compact binary encounter. \\
nsns mergers eject close to one percent of a solar mass  
with an extremely low electron fraction, $Y_e\approx 0.03$. The exact amount of ejecta and their velocities 
depend on both the total mass and the mass ratio of the binary system. Our reference system, the merger of a 
1.3 and 1.4 \Msun binary system, ejects $1.4 \times 10^{-2}$ \Msun at an average velocity of 0.12c. Robust 
r-process within the ejecta produces elements in close-to-solar  proportions from the second to the third 
r-process peak. This abundance pattern is extremely robust and essentially the same
for all investigated cases, see \cite{korobkin12a}. The ejected amounts are consistent with nsns mergers
being a major source of heavy r-process. A question that needs further investigation, though, is 
whether/under which conditions this is consistent with galactic chemical evolution. Earlier work 
\citep{argast04} concluded that neutron star mergers could not be the major source of r-process, 
but all existing studies found that the conditions in merger ejecta are very favorable for r-process 
isotopes to be forged in close-to-solar proportions \citep{freiburghaus99b,goriely11a,roberts11,korobkin12a}
and this discrepancy remains to be understood.
Dynamical collisions as they are expected to occur, for example, in the cores of globular clusters, eject substantially
larger amounts of matter and the overproduction of r-process material by collisions can only be avoided 
if they are rare in comparison to nsns mergers (less, possibly much less, than 10\%).\\
We have also discussed the implications for macronovae, radioactively powered, fast EM transients and for the
radio flares that are produced when the ejecta share their kinetic energy with the ambient medium.\\
The presented amounts of ejecta and their velocities are numerically converged and very robust. They may depend, however, on
the employed physics. In particular, they may be affected by general relativistic and possibly nuclear equation 
of state effects. We suspect that such effects may change the ejecta masses at maximum by a factor of a few, therefore,
the conclusions with respect to the role of mergers as r-process production sites should be robust. 
The sensitivity to GR and nuclear equation of state effects will be explored in future studies. \\
\\
{\bf Acknowledgements}\\
It is a pleasure to acknowledge insightful discussions with 
A. Arcones, E. Berger, A. MacFadyen, O. Korobkin, B. Metzger, E. Nakar, T. Piran, E. Ramirez-Ruiz,
F.-K. Thielemann and I. Zalamea.
Special thanks to E. Gafton who helped preparing Fig.1. This work was supported by
DFG grant RO-3399, AOBJ- 584282. The simulations were performed on the facilities 
of the H\"ochstleistungsrechenzentrum Nord (HLRN), we used SPLASH developed by D. Price
to visualize the hydrodynamics simulations.\\
\\
\noindent Movies from our hydrodynamic simulations and ejecta trajectories can be downloaded from: \\
http://compact-merger.astro.su.se/

\bibliography{astro_SKR}
\bibliographystyle{mn2e.bst}
\bsp

\label{lastpage}

\end{document}